# Equilibrium and out-of-equilibrium dynamics in a molecular layer of azopolymer floating on water studied by Interfacial Shear Rheology

**Davide Orsi, Luigi Cristofolini, Marco P. Fontana**

Physics Department, University of Parma, Viale Usberti 7/A, Parma 43100, Italy

## *Abstract:*

We report the details of the construction and calibration of an ultra sensitive surface rheometer, inspired by the setup described in [C.F. Brooks et al Langmuir 15, 2450 (1999)], which makes use of high resolution video tracking of the motion of a floating magnetized needle and is capable of measuring the viscoelastic response of a Langmuir monolayer with an accuracy of $10^{-5}$ N/m. This instrument is then employed for the rheological characterization of a Langmuir monolayer of a photosensitive azobenzene polymer, which can be brought out of equilibrium by a suitable photoperturbation. The complex dynamic shear modulus $G = G' + i\, G''$ is measured as a function of temperature and illumination power and wavelength. The reversible rheological changes induced in the film by photo-perturbation are monitored during time, observing a transition from a predominantly elastic ($G' > G''$) to a viscoelastic ($G' \approx G''$) regime. These results are confirmed by comparison with independent measurements performed by us using other rheological techniques. Finally a discussion is made, taking into account the results of a recent x-ray photon correlation spectroscopy experiment on the same polymer in equilibrium and out of equilibrium.

## 1. Introduction

Dynamics of molecules confined as thin films is often quite different than in the bulk. For example it is well known that polymer dynamics can be accelerated or reduced in thin films depending on the kind of their interaction with the substrate, resulting in increase or reduction of the glass transition temperature with respect to the bulk value: maximization of the interaction of poly-(2)-vinyl pyridine with a polar substrate results in a 40 °C increase of $Tg$ [1], while a decrease of $Tg$ is observed in thin free standing polystyrene films [2] and polystyrene on silicon [3]. This qualitative finding was quantified by Nealey and co-workers who measured Tg in thin films of polystyrene and poly(methyl methacrylate) on surfaces with different interfacial energies finding a linear scaling of the variation of Tg with the surface energy, irrespective of the chemistry of the polymer [4]. The photosensitivity of an azobenzene-containing polymer was also exploited by us to drive the system out of equilibrium and to study its back to equilibrium relaxation, finding a marked effect of confinement below a threshold thickness of five molecular layers [5]. A comprehensive review and discussion of these effects can be found e.g. in [6]. Furthermore, in some cases the effect of the surface was recently reported to extend even to large depths, on the micron scale, being responsible of an apparently elastic response in micron-thick films of glycerol well above its glass transition temperature [7].

If the situation is more or less well understood for the case of thin films deposited on solid surfaces, less is known for the case of films adsorbed at the air/water interface. However, a great number of practical problems in surface and interface science deal with molecular motions and relaxation at fluid surfaces. And if the physics of molecular layers on solid substrates is made intriguing by the interaction with the substrate itself, the presence of a fluid confining interface makes the problem even more complex, because of the flexibility of the interface (resulting e.g. in the presence of capillary waves which deform the film and are in turn altered by the presence of the film itself) and of the possibility of ions and macromolecules diffusing in the fluid subphase and forming a molecular layer adsorbed to the interface which might alter the properties of the film, as it is the case e.g. for molecular layers of polyaniline on aqueous solutions of salts [8]. In fact, modern soft-matter physics has opened a great number of questions dealing with the dynamics of soft surfaces, and, particularly, with polymer dynamics at interfaces or at reduced dimensionality, e.g. polymer chains in pores or confined between solid walls. Among these systems, adsorbed polymer films represent a paradigmatic example; in fact, polymer films are, frequently, considered as model systems to explore this new physics [9]. Furthermore understanding flow and dynamical transitions in the simple geometry of Langmuir monolayers may be useful in the practical design of cosmetics and food products with desirable textures, in drug delivery, in polymer processing and oil recovery, and in the production of specialized fabric.

Last but not least, rheology of molecular layers is investigated as a means of testing of the fluctuation/dissipation relations -in equilibrium and out-of-equilibrium- in the small systems represented by the Langmuir monolayers, because fluctuations in small systems are larger. For macroscopic systems only the mean values of extensive quantities are of interest because fluctuations around the mean value are ordinarily completely negligible. However in small systems, fluctuations are larger and hence higher moments of the probability distributions become of interest.

At present, there exists no general theory about non-equilibrium systems. Onsager theory and the related fluctuation-dissipation theorems (FDT) prompted probably the most successful attempts: however FDT domain of validity is restricted to the linear response regime [10]. In small systems the situation seems to be the different. Over the past years, a set of theoretical results, that go under the name of fluctuation theorems (FT), have been unveiled [11]. These theorems make specific predictions about fluctuation processes in small systems that can be checked against the behaviour of the small systems constituted by the Langmuir monolayers. In particular, FTs address the microscopic foundations of irreversibility in small systems being central to the puzzle of how time-reversible microscopic equations of mechanics lead to the time-irreversible macroscopic equations of thermodynamics and Langmuir monolayers, as small systems, can be considered at midway between a dynamical system and a thermodynamical one.

In this paper we report our study of the rheological behaviour of a photosensitive azobenzene polymer confined as a Langmuir monolayer. Azobenzene conformation can be changed at will simply by UV and visible light illumination and is responsible for a complex photomechanical behaviour. The phenomenology includes molecular photo-orientation, photo-expansion, and as more recently discovered, photoinduced changes in the viscoelastic response. The use of photoisomerization to produce actuator materials was first proposed by de Gennes [12], and since then a wide literature flourished even though very few real applications emerged. Photoinduced effects on polymeric viscoelasticity are poorly understood at present but their understanding could be the key for a number of applications in several fields, including specialized fabrics, drug delivery and so on. A recent work by Ketner et al. [13] revealed the remarkable potential of aqueous micellar solutions, the viscosity of which can be reduced by more than four orders of magnitude upon UV exposure. The limiting factor in that study was the irreversible character of the transformation; however, in a very recent study [14] the authors demonstrated reversible rheological property changes in an azo-containing siloxane polymer backbone. Such properties are of particular interest for light-activated damping mechanisms, actuable armour, and related applications in fields such as robotics and sensors.

The rest of the paper is organized as follows: we first describe the design of a highly sensitive interfacial shear rheometer and its calibration, we then characterize the different dynamical regimes of a Langmuir monolayer of a photosensitive azopolymer in equilibrium and subject to photoperturbation, and we conclude the manuscript with a discussion of these results also in view of a recent x-ray photon correlation spectroscopy experiment on the same polymer.

## *2. The Interfacial Shear Rheometer*

### 2.1 Design of the instrument

The Interfacial Shear Rheometer we developed is an upgraded version of the instrument described by Fuller [15, 16]. In this instrument, a stress $\sigma(\omega)$ is induced on the monolayer by the oscillating motion of a magnetic needle floating on the surface and driven by the magnetic field gradient generated by a system of coils. The shear $\gamma(\omega)$, along with its phase shift $\delta(\omega)$ respect to $\sigma$, is measured tracking the small needle oscillation with a microscope. The dynamic modulus is given by:

$$G = \sigma(\omega)/\gamma(\omega) e^{i\delta(\omega)} \qquad (1)$$

The imaginary part G' and the real part G'' of the dynamic shear modulus express respectively the elastic and viscous response of the film. In case of predominantly viscous systems, the generalized 2D viscosity $\eta_{2D}$ is obtained from G through the simple relation:

$$\eta_{2D} = \frac{iG}{\omega} \qquad (2)$$

When the viscous response G'' predominates, $\eta_{2D}$ is mainly a real quantity.

The probe used in this experiment is a stainless steel needle (12mm long, 0.34mm thick) whose surface is kept clean by immersion in chloroform. The needle is magnetized to saturation value before each experiment with a permanent magnet. Two coils in Helmholtz configuration (R=16cm), placed around the Langmuir trough (figure 1, left and centre panels), generate a constant field that aligns the needle along their axis. Adding a sinusoidal current to one coil produces the oscillating field gradient that moves the needle. The resulting displacement of the needle is detected by a fast CCD USB camera (The Imaging Source, 1024x768 pixels) equipped with a long focal Mitutoyo MPLAN-APO-20X objective.

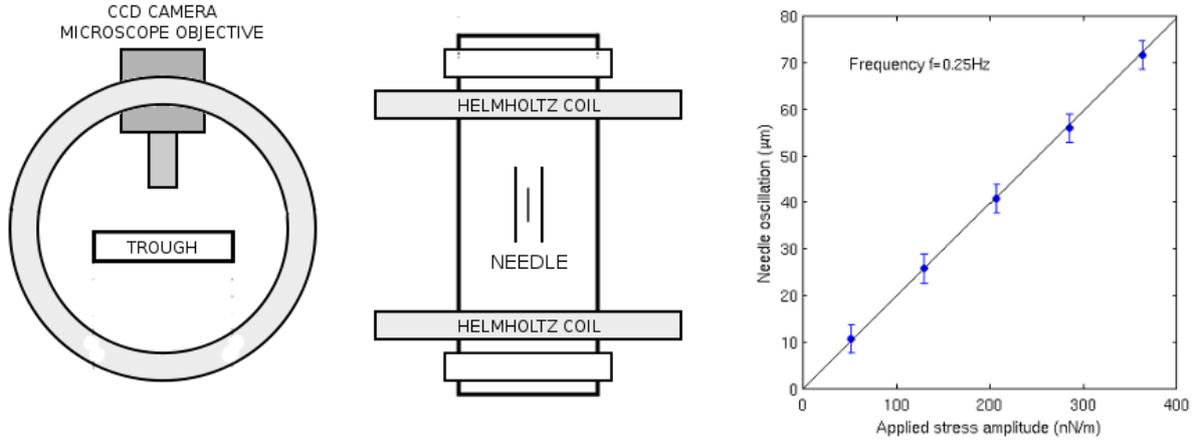

*Figure 1: **left and centre**: simplified scheme of the experimental setup (side and top views); **right**: linearity of the measured displacement versus the applied force, measured on pure water.*

A typical sinusoidal wave of frequency 0.25Hz and amplitude I=1mA generates a force of the order of 50nN on the needle, resulting in a displacement of about 10μm (figure 1, right panel). An hardware DAQ is used to control the instrument: using its analog outputs it is possible to drive the trough barriers and to apply the oscillating current to the Helmholtz coils via a power supply. The DAQ analog inputs are used to measure the trough area, the surface pressure and the current on the coils as function of time.

In order to characterize the inertial response of the needle, a measurement on pure water surface as a function of frequency was performed, in the absence of any surface film. The results could be modelled as the response of a forced-damped oscillator of mass *m*, spring constant *k* and damping constant *d*, in analogy with [16].

$$\frac{\gamma}{\sigma} = \sqrt{\frac{1}{\left(\frac{k}{m} + \omega^2\right)^2 + \left(\frac{\omega d}{m}\right)^2}} \qquad \delta = \arctan\left(\frac{-\omega d}{k - m\omega^2}\right) \qquad (3)$$

The oscillator parameters were determined by a fit performed on the stress-stain ratio curve, as shown in figure 2, left panel. The phase lag $\delta$ is in very good agreement with the model curve built from the very same parameters (figure 2, left panel). Analysing the frequency-dependence of the shear-stress ratio, we obtain that at high frequency the stress-stain ratio is proportional to $\omega^2$ via the inverse of the parameter *m*, which has to be the mass of the needle, the only moving part of the system.

$$\frac{\gamma(\omega)}{\sigma(\omega)} \approx \frac{\omega^2}{m}, \omega \to \infty \qquad (4)$$

Therefore, a current-to-force calibration constant is obtained [15].

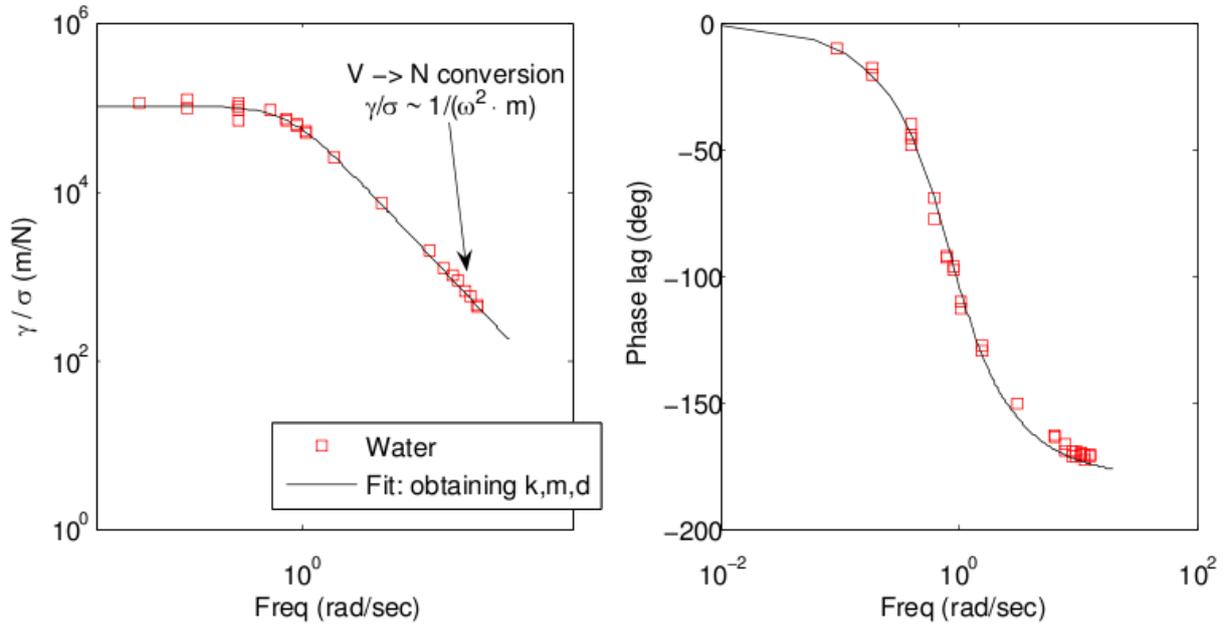

*Figure 2: Inertial response of the needle in absence of a Langmuir layer **left** stress-strain ratio as a function of frequency, fitted with the forced damped oscillator model (eq. 3); **right** oscillations phase lag compared with the result expected for the forced-damped oscillator.*

For the optical tracking of the needle displacements, illumination of the surface trough is necessary. To minimize disturbance to the photosensitive azobenzene material, red light lamp ($\lambda > 550$nm) was employed since the azopolymer has negligible absorption in this frequency region for both its isomers, as shown in the right panel of figure 5. Measurements indicated as performed in dark actually employed this red light background illumination. On the contrary, photoperturbation is obtained using two monochromatic high-power led arrays at 395nm (UV) and 470nm (Blue) from Roithner LaserTechnik, in correspondence of the absorption peaks for the *trans* and *cis* isomers respectively (figure 5, right panel). All the lamps are placed at the centre of the Langmuir trough and produce a wide-angle light cone, to assure the most uniform illumination conditions over the whole film surface. An aluminium box is placed around the whole experimental setup to prevent influences from any external light source.

## 2.2 Rheometer calibration

Prior to its use, the instrument was calibrated measuring the response of films of controlled thickness of poly(dimethylsiloxane) (Aldrich CAS 63148-62-9) of known viscosity, $\eta_{bulk}=0.97$ Pa·s.

Data are shown in the left panel of figure 3: the film behaves as a Newtonian fluid with G'' (circles) growing linearly with the frequency, as indicated by the continuous line, while G' is zero within the error. As a matter of fact, the value found for G' – 4µN/m - can be taken as an estimate of the overall accuracy achievable with this setup. In the right panel of figure 3 we compare the value of the viscosity $\eta_{2D}$ obtained from G'' for each film thickness d with its expected value based on the bulk viscosity:

$$\eta_{2D} = d \cdot \eta_{bulk} \qquad (5)$$

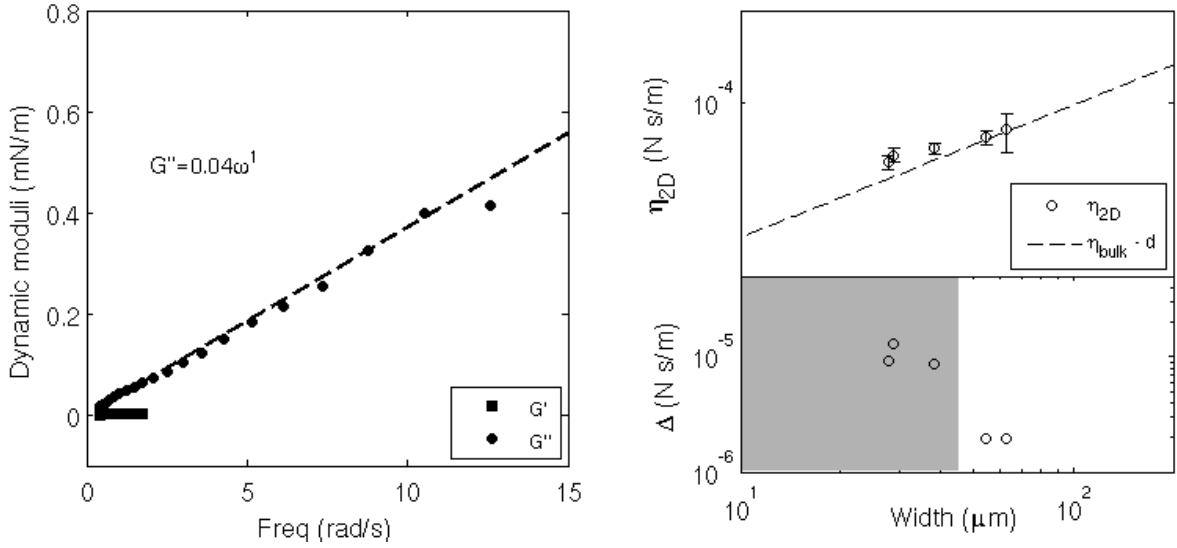

*Figure 3*: *calibration measure on a film of a oil of known viscosity.* **left panel**: *G' is zero within the errors, while G'' is linear in frequency (continuous line), as expected for a Newtonian fluid* **top right panel**: *film viscosity $\eta_{film}$ compared with the expected value $\eta_{bulk}\,d$* **bottom right panel**: *deviation $\Delta=|\eta_{film}-\eta_{bulk}\,d|$, as a function of film thickness d.*

The results are consistent and therefore our instrument can be considered calibrated. Careful inspection of the data from the lowest thickness (d<50μm) however shows a small deviation which can be ascribed to the drag due to the water subphase, which becomes comparable to that of the film when this becomes too thin. This effect has already been found and discussed in the literature [16]. If we suppose, in a simple approximation, a linear superposition of these effects, a rough estimate of the subphase drag is obtained: $|G|_{sub}=10\mu N/m$, a value fully consistent with the accuracy previously obtained from the analysis of G'.

To provide a comparison with independently acquired data, and to validate our results on the azopolymer, we also performed measurements by using a commercial automatic oscillatory ring apparatus (Camtel CIR 100, Camtel Ltd., Royston, U.K.) at the laboratory of Dr Pietro Cicuta in Cambridge University. The instrument was operated at the frequency 2 Hz and amplitude 12mrad.

Finally, some rheological data were also obtained, at least for the azopolymer in its *trans* configuration, by the oscillating barrier technique [17] which relies on the measurement of anisotropy in the surface pressure: the tension is measured as a function of time in orientations perpendicular and parallel to the compression direction and in favourable cases both compression and shear moduli can be deduced.

## 3. Rheological characterization of a photosensitive azopolymer

The calibrated Rheometer was then employed to perform measurements on a particular polymer - poly[[4-pentiloxy-3'-methyl-4'-(6-acryloxyexyloxy)] azobenzene] henceforth called PA4, which has bulk Tg = 20°C and a nematic phase with clearing point $T_{NI}$=92°C [18], Mw=19000 and whose optical properties are shown in the right panel of figure 4: the *trans* isomer is characterized by a strong absorption peak at wavelength 360nm, while the *cis* isomer has little absorption at this wavelength. This material has been already investigated by us both in bulk [19-21] and as molecular layers [5, 22].

More recently we also performed XPCS experiments on this system measuring its microscopic dynamics [23] finding an overall picture consistent with what is found for other polymers and arrested states, e.g. the relaxation time $\tau$ is found inversely proportional to the exchanged momentum, the generalized susceptibility shows a peak. What is peculiar of this photosensitive system is that we also found a signature of photoinduced acceleration of the dynamics indicated by reduction of $\tau$. We note however that XPCS is sensitive to the microscopic space scale, whereas rheological measurements are performed on the macro scale. Extrapolation of rheological properties from XPCS data is therefore a controversial point [24]. With the present work we aim to test this issue.

We first measured the viscoelastic response of the azopolymer in stationary conditions: fixed temperature and illumination. The needle was placed on the water surface at zero pressure, i.e. before starting the compression. The monolayer was then compressed along the isotherm (figure 4) in absence of illumination, turning on the UV light, when required, only at the beginning of the measurement.

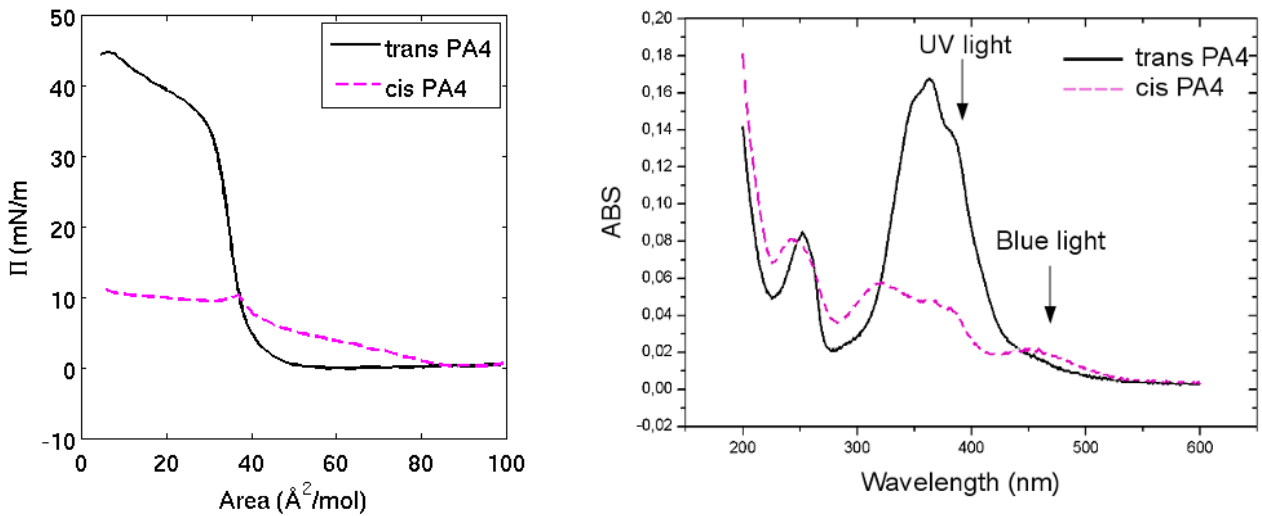

***Figure 4 left***: *compression isotherms of PA4 at T=19°C, in dark and under UV illuminations. Note that the two curves intersect at Π=10mN/m;* ***right***: *absorption spectra for PA4 in dark (trans isomer) and under UV illumination (cis isomer)*

In the same figure 4 we show both isotherms measured in dark and under UV illumination, which are quite different: the *trans* isomer is characterized by a quite steep increase of Π corresponding to the theta solvent situation [22] in the analysis introduced by Vilanove [25] based on the scaling arguments of De Gennes [26]. At the same time the *cis* isomer gives rise to a much softer isotherm, corresponding to favourable interaction of the polymer with the solvent. In order to be able to switch from one isomer to the other at the same pressure and area per molecule, so as to compare rheological behaviours, we choose to perform all our measurements at constant surface pressure Π=10mN/m at which the two curves intersect.

We then performed the rheological measurements, whose results are summarized in figure 5. Based on the linearity test reported in figure 1, we choose to work at constant strain amplitude 10μm, which corresponds to relative strain 0.1%.

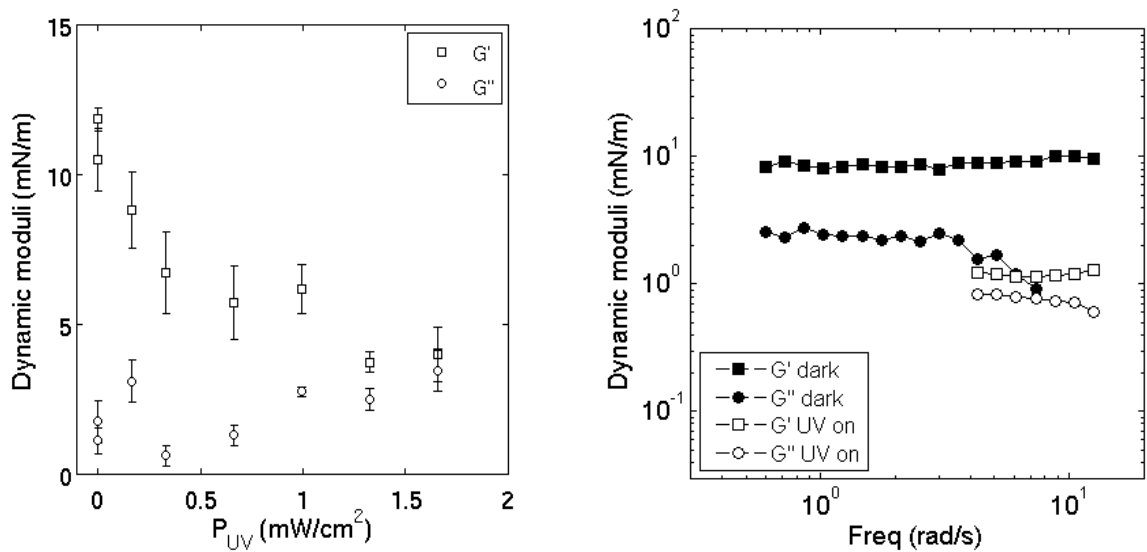

*Figure 5*: *left*: Effect of UV power density on the dynamic shear modulus of a PA4 film measured at 1Hz, at T=22°C and with light source to sample distance of 7.5 cm. Note that a stationary state is reached above 1.3 mW/cm$^2$. *right*: PA4 rheological response at 26 °C and Π=10 mN/m, in dark and under UV illumination. The UV led was placed at 15 cm from the film surface.

First of all, given the possibility of thermally induced spontaneous *cis*-to-*trans* back isomerization of azobenzene, we checked the rheological response of the film at different UV illumination power at fixed temperature and source-to-sample distance l=7.5 cm (figure 6, left panel). A stationary state is reached for irradiation power larger than 1.3mW/cm$^2$. Therefore, in all subsequent experiments, illumination power was set to a higher value, namely 1.6 mW/cm$^2$ thus ensuring complete isomerization.

In the right panel of figure 5 we report the frequency dependence of the dynamic shear moduli G'(ω) and G''(ω) for both isomers of PA4 at T=26°C. In both cases the film response is clearly non-Newtonian, with weak or absent ω-dependence of both G' and G''. The main effect of UV induced cis-isomerization is a fluidification of the film, with a dramatic reduction of the real part G'. While the film of *trans*-PA4 is mainly elastic with tan δ=5, cis-PA4 forms a truly viscoelastic film with G'≈G'' and tan δ= 1. This is in qualitative agreement with the fluidification effects also found by XPCS.

A similar scenario of photoinduced fluidification is found at all the explored temperatures i.e. in the range 15-40 °C. The accessible temperature range is limited on the cold side by the increased rigidity of the polymer film, which makes it brittle and yields unreliable results, and on the warm side by the water evaporation. Work is now in progress to stabilize the water level so as to neutralize the effect of water evaporation.

Figure 6 is an Arrhenius plot summarizing the temperature evolution of the 2D generalized viscosity |η$_{2D}$| measured by ISR and by other rheological techniques, and plotted together with the XPCS relaxation times measured in a PA4 film both in dark and under UV illumination.

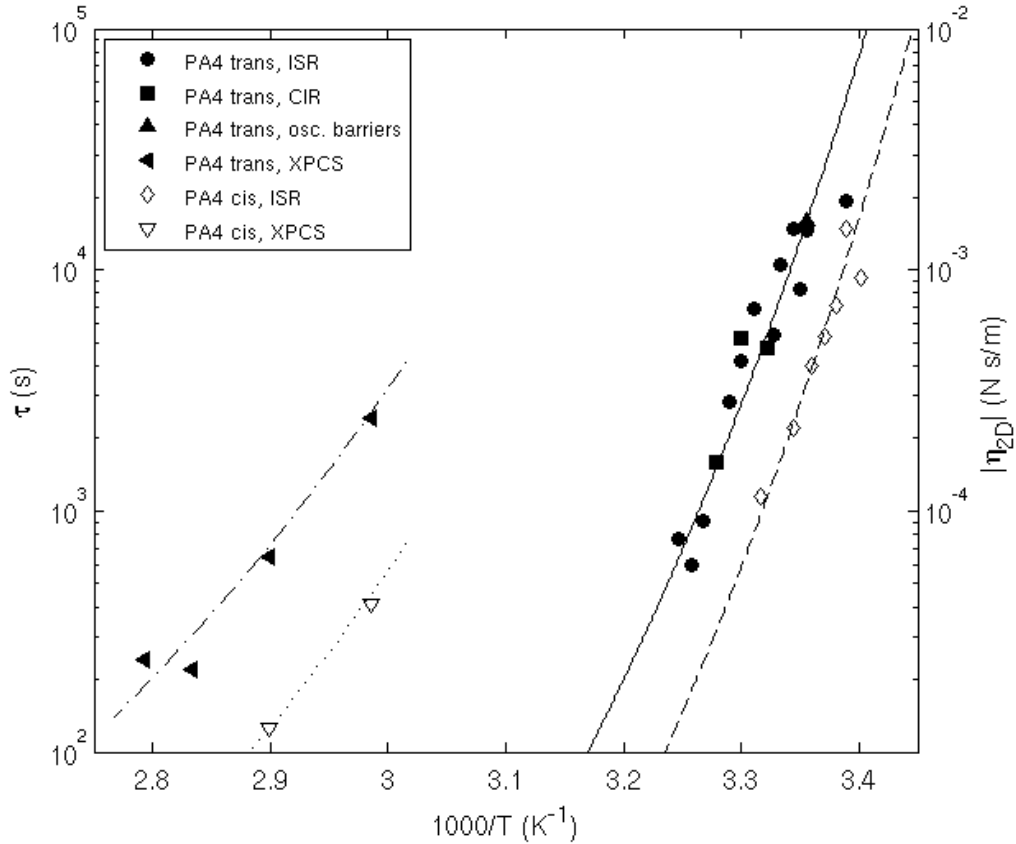

*Figure 6*: Combined Arrhenius plot containing both viscosity $|\eta|$ for PA4 (right y-axis) measured in dark and under UV illumination and the temperature evolution of the XPCS relaxation times in the same conditions. For the viscosity, results obtained with different techniques are reported: ISR, oscillating barriers, and CIR. Filled symbols are used for data measured in dark, while empty symbols stand for measurements under UV illumination. The lines represent the VFT laws describing the temperature dependence, extracted from the bulk viscosity of PA4 (see text for the details) with different pre-exponential factors taking into account the different experimentally measured quantities.

Experimental data are compared with the Vogel-Fulcher-Tammann (VFT) law describing the viscosity of bulk PA4 in dark which was obtained by EPR ($T_0$ =-33 °C, activation temperature $T_A$=997 °C) [27] and was confirmed by depolarized micro Raman and Quartz Crystal Microbalance experiments [21]. Inspection of Fig 6 shows that the same VFT law is observed for both *trans* and *cis* PA4, thus suggesting that it is the main chain dynamics to govern the temperature evolution of the viscosity, while the azobenzene side chains provide an extra hindrance to the motion, which is dependent only on the shape of the isomer and not on the temperature. Furthermore, the fact that XPCS relaxation times, at least for the *trans* isomer, follow the same law as the viscosity, even if with a different pre-exponential factor, implies at least a qualitative agreement between the two techniques as in principle it can be expected based on the Einstein relationship, and which however is known to be a much controversial point [24].

For comparison, and to validate our ISR data, in the same figure we plot the viscosity data obtained by us using other rheological techniques on *trans* PA4, namely the Camtel CIR oscillating ring and by using the oscillating barriers technique. Those data lay on the same trend line as those obtained by ISR, thus confirming our results, which however span a much larger temperature range. The advantage of ISR compared to the oscillating barriers is that ISR allows measurements at multiple frequencies and on a wider range, typically up to 10 rad/sec. In principle the CIR instrument accesses the same frequency range as ISR, however, in its case, to provide well defined boundary conditions one has to confine the film inside a close surface ring, which poses severe problems of film homogeneity and uniform pressure distribution in the case of the rigid films formed by PA4 as by many other polymers. By contrast confinement in ISR is provided by a channel which is open in the direction in which the film is compressed by the Langmuir trough barriers.

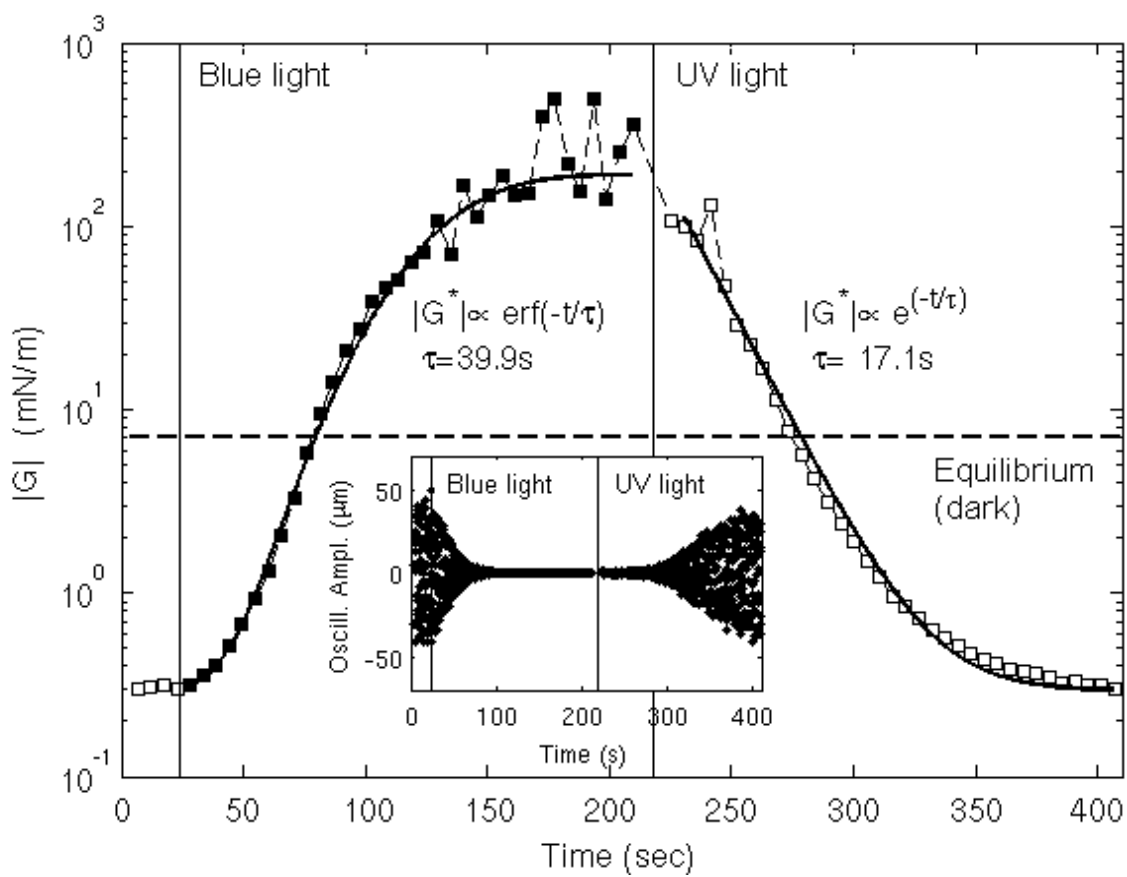

*Figure 7*: *photoinduced transitions in PA4 at 26 °C; |G| is monitored at 1.5 Hz while blue and UV light are alternatively turned on.* **Inset**: *oscillation amplitude as a function of illumination conditions.*

In figure 7 we report the out-of-equilibrium rheological properties of the PA4 Langmuir film measured with variable illumination conditions at constant frequency (1.5 Hz), and T=26 °C. Starting from a thermally stable film formed in absence of light and surface pressure $\Pi$=10mN/m, the system is illuminated alternatively with blue and UV light. A completely reversible change of almost three order of magnitudes in the dynamic shear modulus is obtained through exposure to light. The dynamic shear modulus |G| shows a reversible switching from the more fluid *cis*-PA4 film, reached under UV illumination, to the *trans*-PA4 film induced by blue light. This change in shear modulus is in *qualitative* agreement with the evolution of XPCS relaxation times in PA4 films as reported in [23]. In the same paper we also analysed the out of equilibrium dynamics finding quantitative agreement on the time scales needed for the system to reach equilibrium after photoperturbation, which incidentally agree also with the time scale needed for molecular reorganization as probed by pump-probe photoinduced birefringence measurements [19].

We note in passing that the film under blue-light is characterized by a larger |G| than the all *trans* equilibrium film. This can be explained considering that blue light has two effects on the azobenzene film: on one side it induces *cis* to *trans* isomerization, thus reverting the effects of UV

illumination, on the other side it also induces resonant isomerization, due to the fact that both *trans and cis* isomers have a comparable absorption at this wavelength, as shown in the right panel of figure 4. Therefore resonant isomerization is reasonably responsible of enhancing azobenzene packing thus inducing a more compact and stiff phase than that obtained in the *all trans* film.

Looking in detail at the kinetics of the transitions, we find that both *cis* to *trans* and *trans* to *cis* happen in several tens of seconds (namely 40 and 17 sec respectively), a time scale much longer than that of the photo-switching of the azobenzene moiety [28], reflecting the mesoscopic rearrangements occurring in the film structure and responsible for the very different mechanical properties. The different shape of the two transformations can be explained as follows: in the fluid *cis* film the presence of a small number of *trans* isomers has a negligible effect, the modulus starts to build only when the *trans* concentration grows above some threshold. In the stiff *trans* film on the contrary, even a small concentration of *cis* molecules behaves like a plasticizer thus being effective in reducing the modulus.

## *4. Conclusions*

In conclusion, we have described the construction of a highly sensitive interface rheometer optimized for characterizing a photosensitive azopolymer. The instrument was calibrated and validated on a standard sample constituted by a well characterized commercial oil, and was then employed to study the very different dynamical regimes of a photosensitive polymer, which can be switched between an elastic and a viscoelastic state differing by more than 2 orders of magnitude in their shear modulus. Not only, but also the transient states can be studied, resulting in a slow transition, taking place in several tens of seconds, which we speculate is not driven by the single molecule isomerization time, which is by far shorter, but rather by some *in plane* film reorganization process which is obviously much slower. Our findings also agree with the results of a recent x-ray photon correlation experiment which allowed us to measure relaxation times in both *trans* and *cis* phases of the azopolymer on the microscopic space scale. The results from these two different technique qualitatively agree, despite the very different space scale probed and the fact that XPCS is essentially probing the fluctuations of a system out of equilibrium while ISR is measuring its dissipation response, which should in principle be linked by the Einstein relation which however holds strictly only for systems at equilibrium.

## *Bibliography*

[1] J. van Zanten, W. Wallace, and W. Wu, Physical Review E **53**, R2053-R2056 (1996).
[2] J. Forrest, K. Dalnoki-Veress, J. R. Stevens, and J. R. Dutcher, Physical Review Letters **77**, 2002-2005 (1996).
[3] J. L. Keddie, R. A. L. Jones, and R. A. Cory, Europhysics Letters **27**, 59-64 (1994).
[4] D. S. Fryer, R. D. Peters, E. Jun Kim, J. E. Tomaszewski, J. J. de Pablo, P. F. Nealey, C. C. White, and W. Wu, Macromolecules **34**, 5627-5634 (2001).
[5] L. Cristofolini, S. Arisi, and M. P. Fontana, Physical Review Letters **85**, 4912-4915 (2000).
[6] K. Binder, J. Baschnagel, and W. Paul, Progress In Polymer Science **28**, 115-172 (2003).
[7] L. Noirez and P. Baroni, Journal of Molecular Structure (2010).
[8] L. Cristofolini, M. P. Fontana, O Konovalov, T. Berzina, and A. Smerieri, Langmuir **25**, 12429-34 (2009).
[9] F. Monroy, F. Ortega, R. G Rubio, and M. G Velarde, Advances in Colloid and Interface Science **134-135**, 175-89 (2007).
[10] L. Onsager and S. Machlup, Physical Review **91**, 1505-1512 (1953).

[11] D. J. Evans and D. J. Searles, Advances In Physics **51**, 1529-1585 (2002).

[12] P.G. De Gennes, Physics Letters A **28**, 725-726 (1969).

[13] A. M. Ketner, R. Kumar, T. S. Davies, P. W. Elder, and S. R. Raghavan, J. Am. Chem. Soc. **129**, 1553-9 (2007).

[14] E. Verploegen, J. Soulages, M. Kozberg, T. Zhang, G. Mckinley, and P. Hammond, Angew. Chem. Int. Ed. 3494-3498 (2009).

[15] C. F. Brooks, G. G. Fuller, C. W. Frank, and C. R. Robertson, Langmuir **15**, 2450-2459 (1999).

[16] S. Reynaert, C. F. Brooks, P. Moldenaers, J. Vermant, and G. G. Fuller, J. Rheology **52**, 261 (2008).

[17] P. Cicuta and E. M. Terentjev, European Physical Journal. E, **16**, 147-58 (2005).

[18] A. S. Angeloni, D. Caretti, C. Carlini, E. Chiellini, G. Galli, A. Altomare, R. Solaro, and M Laus, Liq. Cryst. **4**, 513 (1989).

[19] P. Camorani and M. P. Fontana, Physical Review E **73**, 11703 (2006).

[20] L. Cristofolini, M. P. Fontana, M Laus, and B Frick, Physical Review E **64**, (2001).

[21] L. Cristofolini, P. Facci, P. Camorani, and M. P. Fontana, J Phys.: Cond Matt **11**, A355-A362 (1999).

[22] L. Cristofolini, M. P. Fontana, T. Berzina, and O. Konovalov, Physical Review E **66**, (2002).

[23] D. Orsi, L. Cristofolini, M. P. Fontana, A. Madsen, and A. Fluerasu, Submitted To Physical Review E (2010).

[24] A. Papagiannopoulos, T. A. Waigh, A. Fluerasu, C. Fernyhough, A. Madsen, J Phy: Cond Matt **17**, L279-L285 (2005).

[25] R. Vilanove and F. Rondelez, Physical Review Letters **45**, 1502-1505 (1980).

[26] P.G. De Gennes, *Scaling Concepts In Polymer Physics* (Cornell Univ Press, 1979).

[27] L. Andreozzi, Ph.D. Thesis, University of Pisa, Italy (1997).

[28] R. Hildebrandt, M. Hegelich, H. Keller, G. Marowsky, S. Hvilsted, C. N. R. Holme, and P. S. Ramanujam, Phys Rev Lett **81**, 5548-5551 (1998).